%% file: main.tex
\documentclass[sigconf]{acmart}
\AtBeginDocument{%
  \providecommand\BibTeX{{%
    \normalfont B\kern-0.5em{\scshape i\kern-0.25em b}\kern-0.8em\TeX}}}

\copyrightyear{2025}
\acmYear{2025}
\setcopyright{cc}
\setcctype{by-sa}
\acmConference[IUI '25]{30th International Conference on Intelligent User Interfaces}{March 24--27, 2025}{Cagliari, Italy}
\acmBooktitle{30th International Conference on Intelligent User Interfaces (IUI '25), March 24--27, 2025, Cagliari, Italy}
\acmDOI{10.1145/3708359.3712115}
\acmISBN{979-8-4007-1306-4/25/03}

\usepackage{nawta}

\begin{document}

\title{Dynamik: Syntactically-Driven Dynamic Font Sizing for Emphasis of Key Information}

\author{Naoto Nishida}
\email{nawta@g.ecc.u-tokyo.ac.jp}
\orcid{0000-0001-9966-4664}
\affiliation{%
  \institution{The University of Tokyo}
  \city{Tokyo}
  \country{Japan}
}

\author{Yoshio Ishiguro}
\email{ishiy@acm.org}
\orcid{0000-0002-1781-6212}
\affiliation{%
  \institution{The University of Tokyo}
  \city{Tokyo}
  \country{Japan}
}

\author{Jun Rekimoto}
\email{rekimoto@acm.org}
\orcid{0000-0002-3629-2514}
\affiliation{%
  \institution{The University of Tokyo}
  \city{Tokyo}
  \country{Japan}
}
\affiliation{%
  \institution{Sony CSL Kyoto}
  \city{Kyoto}
  \country{Japan}
}

\author{Naomi Yamashita}
\email{naomiy@acm.org}
\orcid{0000-0003-0643-6262}
\affiliation{%
  \institution{Kyoto University}
  \city{Kyoto}
  \country{Japan}
}

\renewcommand{\shortauthors}{Naoto Nishida, et al.}

\begin{abstract}
\input{00_abstract}
\end{abstract}



\begin{teaserfigure}
  \includegraphics[width=\textwidth]{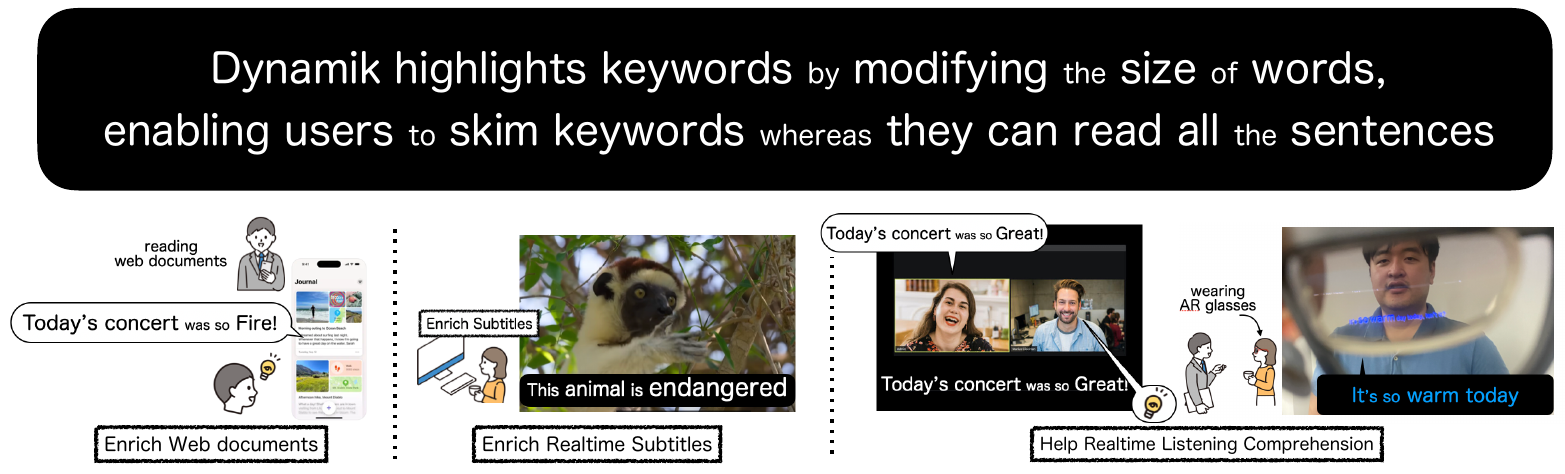}
  \caption{Dynamik is an automated keyword highlighting system for non-native speakers in a specific language when they use audio speech recognition to aid their listening skill. it highlights important keywords by modifying the size of words, enabling users to skim keywords whereas they can read all the sentences.}
  \Description{Dynamik highlights important keywords by modifying the size of words, enabling users to skim keywords whereas they can read all the sentences.}
  \label{fig:teaser}
\end{teaserfigure}


\maketitle

\input{01_introduction}
\input{02_related_work}
\input{04_design}
\input{05_implementation}
\input{06_evaluation}

\input{07_result}

\input{08_discussion}

\input{09_conclusion}

\input{99_recycle}
\begin{acks}
This work was supported by JST BOOST Grant JPMJBS2418, JST Moonshot R\&D Grant JPMJMS2012, JST CREST Grant JPMJCR17A3, and the commissioned research by NICT Japan Grant JPJ012368C02901.
\end{acks}

\bibliographystyle{ACM-Reference-Format}
\bibliography{main}

\appendix
\input{10_appendix}

\end{document}

%% file: 00_abstract.tex
In today's globalized world, there are increasing opportunities for individuals to communicate using a common non-native language (lingua franca). 
Non-native speakers often have opportunities to listen to foreign languages, but may not comprehend them as fully as native speakers do. To aid real-time comprehension, live transcription of subtitles is frequently used in everyday life (e.g., during Zoom conversations, watching YouTube videos, or on social networking sites). However, simultaneously reading subtitles while listening can increase cognitive load. 

In this study, we propose Dynamik, a system that reduces cognitive load during reading by decreasing the size of less important words and enlarging important ones, thereby enhancing sentence contrast. Our results indicate that Dynamik can reduce certain aspects of cognitive load, specifically, participants' perceived performance and effort among individuals with low proficiency in English, as well as enhance the users' sense of comprehension, especially among people with low English ability.
We further discuss our methods' applicability to other languages and potential improvements and further research directions.

%% file: 01_introduction.tex
\section{Introduction}

In today's globalized society, there are more and more situations in which people speak common language (\ie lingua franca) and communicate with each other. 
However, for many people around the world, speaking a language that is not their native tongue can cause difficulties, especially when they listen to native speakers~\cite{Vandergrift_2007}. 
To address this problem, subtitling has become widely used in videoconferencing systems and online video platforms as a listening aid~\cite{Hwang:10.1080/10447318.2018.1543091, abdellah2008can}.
However, listening to audio and reading subtitles at the same time involves performing different tasks simultaneously, further increasing the cognitive load of non-native speakers~\cite{Paas2004CognitiveLT, Kalyuga2011, Mayer2001CognitiveCO}. 
Thus, there is a demand for subtitles that allow users to quickly obtain information on important parts of a text, rather than conventional subtitles~\cite{Caimi2006AudiovisualTA, hausataari2014}.

In past studies, many psychological experiments have been conducted to investigate the effects of various styles of texts (\eg fonts, sizes, etc.) on reader psychology, but there are limited studies that focus on alleviation of cognitive load.
As examples of limited studies, Pan et al. showed that it decreases the burden on non-native users to have native speakers highlight the important parts of the subtitles by touching each word to allow non-native speakers to see the highlighted subtitles when they listen to native speakers~\cite{Pan2017}.
Since native speakers have more leeway to annotate important keywords compared to non-native speakers, native speakers annotate keywords in this research. 
However, the automation of keyword would benefit both because native speakers can focus solely on the topic they are speaking on, and the other behaviors such as gestures or observation of the listeners. 
Hausataari et al. examined whether terminology-highlighted subtitles alleviate non-native speakers' workload necessary for reading, in the case of catching up the conferences' speech~\cite{Hausataari2014_keyword}. 
The authors proved that termiology-highlighted subtitles did not improve the comprehension of nonnative speakers, but since speech recognition accuracy was not so good in that period (Word Error Rate was 23\%), it is assumed that terminology could not be well recognized at the time.

In this study, we propose a system called ``\textit{Dynamik}'' to address the alleviation of users' workload by automation of keyword highlighting in real-time. 
\textit{Dynamik} classifies the importance of words in speech-recognized sentences into content words and function words based on morphological analysis and displays less important words (\ie function words) smaller and more important words (\ie content words) larger in real time, thus displaying the contrast of importance of each word visually to improve readability (Figure. \ref{fig:teaser}). 
This method is expected to reduce the workload of non-native speakers listening and to achieve more effective information transfer.
We conducted a crowd-sourcing experiment with 84 participants to investigate the effectiveness of our method in English. 

The result showed that \textit{Dynamik} was significantly preferred by non-native English speakers but not by native English speakers, that the workload items of \textit{Performance} and \textit{Effort} were significantly lighter and increased the sense of comprehension than under the other conditions such as common subtitles and the subtitles that only show keywords. 

Our method is applicable to languages other than English, and we expect to reduce the display area of subtitles by reducing the size of unimportant words. 

Our contributions are presented below:
\begin{itemize}
    \item We investigated the possible automation design of accessible subtitles for non-native English speakers that works in realtime. To the best of our knowledge, this is the first work to automate the highlighted keyword subtitle from morphological aspects.
    \item We developed and published a real-time dynamic subtitle system which we applied to build \textit{Dynamik}~\footnote{\url{https://github.com/nawta/Dynamik_client}\label{fot:dynamik_client}}~\footnote{\url{https://github.com/nawta/Dynamik_server}\label{fot:dynamik_server}}. This system can be used for other methods to modify subtitles with subtle changes (\eg changing font color, other definition of keywords other than morphoanalysis).
    \item We presented \textit{Dynamik}, a novel subtitling method to assist non-native English speakers during listening, focusing on function words and content words. No prior work used these perspectives for workload alleviation.
    \item We published a web app example of secure psychological experiments for crowd-sourcing by encryption of external files and obfuscation of codes, which we used for our experiment~\footnote{\url{https://github.com/nawta/Dynamik_experiment}\label{fot:dynamik_experiment}}. This code is especially useful for when you have something you don't want participants to read in the projects (in our case, the answer of quizzes and completion codes).
    \item We conducted evaluation experiments with 84 people and showed that \textit{Dynamik} alleviates several workloads (\textit{Effort}, \textit{Performance}) and affords more self-awareness of comprehension for non-native English speakers.
\end{itemize}

%% file: 02_related_work.tex
\section{Related Work}
Our research focuses on alleviating workloads during non-native English speakers' communication.
First, we discuss a communication challenge for non-native English speakers and existing solutions for it.
Second, we discuss existing research methods on media richness for subtitles to deepen our solution.
Third, we discuss several examples for extracting keywords or key phrases to discuss our concrete implementation.
Lastly, we further discuss possible applications of our research to consider our keyword extraction method.

\subsection{Communication Challenge for Non-Native English Speakers}
Non-native English speakers often find themselves compelled to communicate in English, even though they do not have the same level of comprehension as native English speakers. 
This situation creates a significant cognitive load that can be further exacerbated by the provision of full, unfiltered subtitles. 
A potential solution to this problem might involve selective presentation of information, omitting less crucial elements, or highlighting important elements.

Although direct solutions to these issues have not been extensively proposed, related research in the fields of psychology, computer graphics, and computer-aided language learning (CALL) has produced various relevant findings~\cite{Higgins_1983, levy1997, doi:10.1080/15213269.2010.502873, 10.1109/TVCG.2015.2467759}.

In the field of CALL, numerous studies have explored the use of audio and text subtitles for the learning of foreign languages. For example, Hwang compared the effects of fragmented subtitles versus standard subtitles on English learning outcomes and cognitive load~\cite{Hwang:10.1080/10447318.2018.1543091}. The study concluded that fragmented subtitles increased cognitive load and were more effective in English learning.
Another notable approach is \textit{Flash Word}, which provides fragmented audio synchronized subtitles, using keywords extracted through tf-idf to assist English as a Second Language (ESL) learners~\cite{flashword, tfidf}.

It is important to note that while these studies aim to increase cognitive load for learning purposes, our research focuses on reducing the cognitive load associated with reading subtitles during listening tasks.

\subsection{Subtitle Display Methods}

Research on enhancing lean media with rich elements has a long history in the field of media richness and accessibility~\cite{10.1287/mnsc.32.5.554, 10.1111/j.1468-2958.1993.tb00309.x, higgins1996activation, 10.1145/3411764.3445412}.

Regarding subtitling, many studies, particularly those focusing on Deaf and Hard of Hearing (DHH) individuals, have explored ways to incorporate non-verbal elements of speech into subtitles~\cite{10.1145/638249.638284, 10.1145/2543578, 10.1145/3613904.3642258}. 
For individuals with dyslexia, some proposed ways to facilitate text skimming~\cite{Turney2002, Rello2015HowTP}.
These studies have examined various aspects of text presentation, including font size~\cite{7576622, 10.1145/642611.642677}, height\cite{10.1145/3544548.3581130}, highlighting~\cite{10.1145/3308561.3353781}, font color~\cite{PONCE201421, 10.1145/3132525.3132541}, typeface change~\cite{10.1145/2700648.2809843}, font weight~\cite{Turney2002, rello-etal-2014-keyword}, appending emojis~\cite{10.1145/1279540.1279551, 10.1145/3373625.3418032}, upper case~\cite{berger2018}, underlining~\cite{10.1145/1357054.1357288}, bold or italic~\cite{rello-etal-2014-keyword, berger2018}, transparency~\cite{6297574}, text spacing~\cite{nanbo2012}
, syllables~\cite{alina2011}, removal of text~\cite{nagasawa2013}, dynamic positioning~\cite{10.1145/2636242.2636246,10.1145/1873951.1874013}.
In addition, research in cognitive psychology has investigated how font size and other typographic elements influence text perception~\cite{https://doi.org/10.1111/j.2044-8317.1955.tb00160.x, wehr2004typography, https://doi.org/10.1111/j.2044-8295.1989.tb02317.x}.

In the field of text design, principles have been established showing that variation in the emphasis of text can improve the comprehension of ideas~\cite{principlesofdesign}. This principle suggests that differentiating between keywords and non-keywords through visual emphasis could promote understanding.

Our research targets non-native speakers, and we have drawn inspiration from these previous studies on subtitle variations to create a design inspiration for our implementation.

\subsection{Keyword Extraction Method from Linguistic Perspective}

Several approaches to keyword extraction have been proposed in previous research in the field of Natural Language Processing, including statistical methods (derived from tf-idf)~\cite{tfidf, 10.1561:1500000019, Zaragoza2004MicrosoftCA, 10.1145/1031171.1031181, Lv2011LowerboundingTF, CAMPOS2020257, el-beltagy-rafea-2010-kp}, graph-based methods (derived from TextRank)~\cite{mihalcea-tarau-2004-textrank, wan-xiao-2008-collabrank, bougouin-etal-2013-topicrank, Sterckx2015TopicalWI, florescu-caragea-2017-positionrank, boudin2018unsupervisedkeyphraseextractionmultipartite}, and some Seq2seq neural network models (derived from word embedding)~\cite{grootendorst2020keybert, Schopf_2022, Sharma2019SelfSupervisedCK}.
However, most of these are supposed to be used a posteriori to preexisting transcripts. Seq2seq models can predict a priori, but still require training.
To measure our Proof-of-Concept quickly, we first looked for a method that could be done without training.

From a linguistic perspective, content words (nouns, main verbs, adjectives, adverbs) carry specific meanings, while function words serve primarily grammatical roles and contribute less to the content of the sentence~\cite{fries1952, Carnap1937-CARTLS-4}.
In English speech, content words are typically stressed, while function words are unstrained, reflecting their relative contribution to information content~\cite{10.1037/0033-295x.99.2.349}.

Based on the aforementioned design principles and these findings, if content words are defined as keywords, it may be possible to promote comprehension by making content words more prominent in sentences while making function words less prominent.

Therefore, \textit{Dynamik}, the system we propose in this study, integrates these findings by displaying function words in smaller text and content words in larger text, aiming to support non-native English speakers' listening comprehension.

\subsection{Subtitle Display Area}

Research on subtitle display areas and placement has been conducted primarily in the context of visualization studies~\cite{10.1145/3613904.3642772, gerber2020effects}. Minimizing the subtitle display area has been a goal in these studies, with potential benefits for devices with limited display areas, such as smart glasses.

The proportion of function words to content words in a text is referred to as Lexical Density, defined by the following formula by Halliday~\cite{halliday_1985}:

\begin{displaymath}
\rm{Lexical\,Density} = \frac{\rm{Number\,of\,Content\,Words}}{\rm{Total\,Number\,of\,Clauses}} * 100
\end{displaymath}

Typically, function words account for approximately 40\% of the words in a text~\cite{klammer2009}. By reducing the size of or omitting function words, we can potentially decrease the overall subtitle display area.

As discussed in our design section later, we chose to visually de-emphasize function words and emphasize content words to support non-native English speakers' listening comprehension while potentially reducing the subtitle display area.

%% file: 04_design.tex
\section{Design}

\subsection{Core Concept}

The central idea of this research is to enhance non-native English listeners' ability to efficiently extract information from text by emphasizing keywords and deemphasizing less crucial words. 
In short, our goal is to develop subtitles that facilitate more effective skimming.

\subsection{Defining Keywords}
The definition of content words (\ie \textit{words that possess semantic content and contribute to the meaning of the sentence in which they occur}) and function words (\ie \textit{words that have little lexical meaning or ambiguous meaning and express grammatical relationships among other words within a sentence}) has shown that the classification of words into content words and function words strongly correlates with their contribution to the overall meaning of a text~\cite{fries1952}. 
As textual content positively correlates with information density, we can broadly categorize content words as essential and function words as less critical for our purposes.
In addition, in phonology, the pronunciation patterns of words in speech can also inform this classification (\ie importance). Generally, function words are pronounced less prominently than content words~\cite{194c8072-4e06-31a3-9bcc-16858153f7f2}. 
Taking these factors into account, we have designated the following parts-of-speech as important ``keyword'':

\begin{itemize}
    \item Nouns
    \item Verbs
    \item Adjectives
    \item Adverbs
    \item Negatives
\end{itemize}

\subsection{Subtitle Presentation Method}

\fgw{experiment_window_v3}{experiment_window_v3}{1.0}{Three conditions in the experiment: \textit{Normal} subtitle (left), \textit{Keyword} subtitle, which reduces the font size of function words (center), \textit{Dynamik} subtitle, which reduces the font size of function words (right).}

Drawing from text design principles and variations in prior research on subtitles, we held a one-hour discussion session on several potential approaches for presenting assistive subtitles. 
The authors and one invited Natural Language Processing researcher participated in this session.

Regarding the color used in the experiment, we chose bright pink ((R, G, B) = (225, 128, 130)) for the texts and black for the background of Color Universal Design~\cite{universalcolor} because the previous study found that the combination of creme color and black is easy to read ~\cite{Rello_2017} and the dark black background is often used for AR systems.

Considering the potential applications of our system, we envisioned its use not only on computer monitors but also on devices where screen real estate is at a premium, such as Head-Mounted Displays (HMDs) and emerging smart glasses. 
In addition, there are options that are not suitable for all fonts (\eg bold, italic), leading to our method's lack of generalizability.
Given this context, we prioritized minimization of screen occupancy in our design approach. Consequently, we chose to implement two methods to compare with normal subtitles (hereafter `\textbf{Normal}') (See \fgref{experiment_window_v3}):
\begin{itemize}
    \item Reducing the font size of function words (hereafter `\textbf{Dynamik}')
    \item Omitting function words entirely (hereafter `\textbf{Keyword}')
\end{itemize}
We named \textit{Dynamik} after the term for musical expression through changes or contrasts in the intensity of the sound.




%% file: 05_implementation.tex
\section{Implementation}

\fgw{Imprementation_Dynamik}{Imprementation_Dynamik}{1.0}{System workflow. It begins with capturing English audio input using the PC's built--in microphone. The audio is then processed through speech recognition, followed by morphological analysis of the real-time speech recognition results. Based on this analysis, words that are not considered content words (nouns, adjectives, verbs, and auxiliary verbs) are reduced in size or omitted, depending on the subtitle type. The processed text is then displayed on a Unity-based interface.}

We developed three types of assistive subtitles for our study: \textit{Dynamik}, which reduces the size of less important words; Keyword, which completely omits less important words entirely; and \textit{Normal}, standard subtitles as a control condition.

The system workflow is shown on \fgref{Imprementation_Dynamik}.
It begins with capturing English audio input using the PC's built--in microphone. The audio is then processed through speech recognition, followed by morphological analysis of the real-time speech recognition results. Based on this analysis, words that are not considered content words (nouns, adjectives, verbs, and auxiliary verbs) are reduced in size or omitted, depending on the subtitle type. The processed text is then displayed on a Unity-based interface. 

For the client-side implementation, we used Unity version 2022.3.21f1 with C\# 12.0 and .NET Framework 8.0~\cite{unity, csharp, dotnet}, while the server-side was implemented in Python 3.10~\cite{python}. 
We used the Azure Speech Recognition API for speech recognition~\cite{azure} and spaCy 3.7.5 with 
en\_core\_web\_sm 3.7.1 for morphological analysis and part-speech tagging~\cite{spacy, encore}. 
Part-of-Speech tagging was performed using a Hidden Markov Model implemented in spaCy, which also provided functionality for stop word detection. 
Communication between the client and the local server was facilitated using ZeroMQ~\cite{zeromq}, and between the client and the Azure server was facilitated using the Azure SDK for .NET~\cite{azure_sdk}.
The codes are available here~\footref{fot:dynamik_client} and here~\footref{fot:dynamik_server}.
Our development and testing were conducted on a 13--inch MacBook Pro (2021 M1 model).

The system refreshes the subtitles every 0.5 seconds. 
Morphological analysis and parts of speech take approximately 0.2--0.3 seconds for English text ( < 0.5 seconds ) on the local server, and the communication between the client and the servers took infinitesimal compared to language processing. 
Therefore, we did not find significant differences in the subtitle presentation intervals between the different methods due to consistent processing times.

For the subtitle size, we used a standard font size of 18\,pt for every condition, except for the case that \textit{Dynamik}'s function words were displayed at 12\,pt. 
These sizes were chosen according to the Web Content Accessibility Guidelines~\cite{wcag}. The guidelines says that ``\textit{with at least 18 point or 14 point bold or font size that would yield equivalent size for Chinese, Japanese, and Korean (CJK) fonts}'' for large scale, and ``\textit{For many mainstream body text fonts, 14 and 18 points are roughly equivalent to 1.2 and 1.5 em or to 120\,\% or 150\,\% of the default size for body text}'' for font jumps.
The sizes are also based on the readability research on dyslexia, which about 10\,\% of people have)~\cite{shaywitz1992evidence,doi:10.1177/002221949302600501}. 
The 18\,pt size offers optimal readability for a broad audience, including those with dyslexia and the elderly~\cite{10.1145/2461121.2461125, 10.1145/2858036.2858204, 10.1145/634067.634173}, while the 12\,pt is the smallest recommended size that maintains readability for the same diverse group~\cite{bhatia2011effect, 10.1145/2207016.2207048}.

For subtitle typefaces, we chose ZenMaruGothic Medium~\cite{zenmarugothic}, because it satisfies ``monospaced, San Selif'' features, which are supposed to be easy to read for both people with and without dyslexia~\cite{10.1145/2897736, 10.1145/2513383.2513447}.
The color of the subtitles was selected as (R,G,B,A)=(255, 128, 130, 255), which is selected from a color palette for visually impared people~\cite{universalcolor}.



%% file: 06_evaluation.tex
\section{Evaluation}

To determine which assistive subtitle method offers the highest readability and reduces cognitive load, we conducted a crowd-sourced experiment involving 104 participants. This section details the participants, the experimental application, the procedure, and its result.

\subsection{Participants}

We used Prolific for crowd-sourcing~\cite{prolific}. Out of 104 initial participants, we excluded two who failed our dummy quiz, one who couldn't complete the experiment within the time limit (67 minutes, as calculated by Prolific), and 17 who didn't finish all responses. This left us with data from 84 participants for analysis. We included 18 native English speakers as a comparison group. We offer a £9.9 / hour -- £12 / hour reward to the participants.

\subsection{Experimental Application}

We developed a custom experimental application using Vue.js 3.2.13~\cite{vue}, Node.js v20.17.0 (npm v10.8.2)~\cite{nodejs, npm}, and Webpack 5.3.10~\cite{webpack}. This setup allowed us to modify the directory structures between development and deployment, improving security. We used jsPsych 7.3.4 to create the experimental workflow~\cite{Leeuw2023} and Webpack Obfuscator 3.5.1 for the obfuscation of code~\cite{webpack}.
The code is available here~\footref{fot:dynamik_experiment}.

For data storage, we used jsPsychSheet~\cite{jsPsychSheet}, DataPipe~\cite{datapipe}, and the Open Science Framework~\cite{osf}. 

For some figures inserted during the experiment for listening comprehension quizzes, we used Adobe Firefly~\cite{adobe_firefly}.

To protect highly credential content, such as question texts and completion codes, we encrypted external files using a combination of the XOR cipher, string concatenation/splitting, and Caesar cipher.

\subsection{Procedure}
\fgw{Experiment_flow}{Experiment_flow}{1.0}{Main part of the experiment workflow. i) Participants listened to an audio track and completed 10 TOEFL-adapted listening comprehension questions (\textit{Pre-test}) to assess their English listening skills. ii) The participants then listened to six CNN news clips~\cite{cnn} with assistive subtitles (either one from \textit{Normal}, \textit{Keyword}, or \textit{Dynamik}). After each excerpt, participants completed three questionnaires to ask their self-awareness of the extent of engagement with watching the clip, their self-awareness of the extent of comprehension of the clip, and the readability of the subtitle during the clip. After that, they also completed NASA--TLX assessments and listening comprehension quizzes (\textit{Comprehension Quiz}) on the clip~\cite{HART1988139}. This step is repeated six times with the randomized order of the video clips (two clips for every condition).}
The main part of the experiment workflow is shown in \fgref{Experiment_flow}.
The detailed experimental workflow was as follows:

\begin{enumerate}
    \item Participants accessed the experiment page through Prolific.
    \item They consented to the experiment and provided basic information (gender, nationality, native language, other spoken languages, and English test scores such as TOEFL~\cite{toefl}, TOEIC~\cite{toeic}, IELTS~\cite{ielts}).
    \item Participants listened to an audio track and completed 10 TOEFL-adapted listening comprehension questions (hereafter `\textbf{Pre-test}') to assess their English listening skills. The audio track and five questions were from the 5th Edition TOEFL Official Guide (track 1)~\cite{officialguide}, and five were generated using Claude 3.5 Sonnet and validated by the authors (\eg so that the questions do not interfere with the others as hints)~\cite{claude}.
    \item The participants then listened to six CNN news clips~\cite{cnn} with assistive subtitles (See \tbref{news_clip_info}). We used three subtitle conditions (\textit{Normal}, \textit{Keyword}, and \textit{Dynamik}), with two excerpts per condition presented in random order. 
    \item After each excerpt, participants completed three questionnaires to ask their self-awareness of the extent of engagement with watching the clip, their self-awareness of the extent of comprehension of the clip, and the readability of the subtitle during the clip.
    After that they also completed NASA--TLX assessments and listening comprehension quizzes (hereafter `\textbf{Comprehension Quiz}') on the clip~\cite{HART1988139}. We incorporated dummy tests with a probability of appearance 40\% on each NASA--TLX and the \textit{Comprehension Quiz} page.
    \item After repeating step (5) six times, the participants commented on their impressions of the experiment to conclude the experiment.
\end{enumerate}

The order of the answer choices for \textit{Pre-test} and \textit{Comprehension Quiz} was randomized. 
We adjusted rewards based on \textit{Comprehension Quiz} performance, offering £3.3 for standard completion (estimated at 20 minutes, £9.9/hour) and £4 for scores above 80\% (£12 / hour), although no participant achieved this threshold.
The NASA--TLX and \textit{Comprehension Quiz} are given six times, which means that the probability of no dummy quiz on each item is $(6/10)^6=0.046... < 0.05$.
The CNN video clips were 40--60 seconds long, with a resolution of $960\,\times\,540$, a video quality of 21\,Mbps and an audio quality of 48.0\,kHz. All kinds of subtitles were updated on the screen every 0.5 seconds.
The average duration of the experiment was 25 minutes and 47 seconds.
We also had a free-form questionnaire at the end of the experiment.

\subsubsection{Additional TOEFL Questions}
Here's the prompt we used to create additional TOEFL questions:
``Please create an additional question for the following TOEFL statement. Make sure that the questions do not cover the same subject matter as the original question. The original question is as follows: \{\textit{Original Questions}\}''

\subsubsection{Quiz Validity}
To assess listening comprehension of the news clip, we generated quizzes (\textit{Comprehension Quiz}) using AI models (Claude 3.5 Sonnet~\cite{claude}, Gemini 1.5 Pro~\cite{gemini}, and ChatGPT 4o~\cite{chatgpt}). 
The prompt we used to generate additional \textit{Comprehension Quiz} was ``Attached are 6 independent CNN news transcriptions. For each of these news items, create three quizzes to test your understanding of the content. The quiz has four choices for each question and only one of the choices is correct''.
After that, we discussed the validity of the quiz questions and the correct answer choices, make revisions of the contradictions, and then reduce the number of questions from nine to seven to ensure that they did not interfere with other questions.
We attach the quizzes and the answer in the appendix.
The information of the \textit{Comprehension Quiz} is shown in the Appendix (See \tbref{quiz_info1}, \tbref{quiz_info2}, \tbref{quiz_info3}, \tbref{quiz_info4}, \tbref{quiz_info5} and \tbref{quiz_info6}).


%% file: 07_result.tex
\section{Result}
Here, we discuss the results of each metric in the experiment.

\subsection{Distribution of Conditions for Each News Clip}

\fgw{video_condition_distribution}{video_condition_distribution}{1.0}{Condition distribution for each video news clip.}

\fgref{video_condition_distribution} shows the condition distribution for each news clip.
Since we assigned each participant to watch every clip, each video was watched exactly 84 times and there was no significant bias on the distribution by conditions.

\subsection{Demographics}
\fgw{demographic_distributions}{demographic_distributions}{0.8}{Demographic distribution of the survey. Note that the color matches the language in graphs A) and B) but not in graphs C) and D). A) Distribution of native languages B) Distribution of other languages that the participants can speak. C) Native language distribution, the same as A), but we modified its color according to the language families. D) Distribution of nationality.}
\fgref{demographic_distributions} shows the demographic data of the participants in this experiment. We recruited participants whose native languages are full of variety (\fgref{demographic_distributions} (A), (D)).
We recruited 18 English native speakers as well as English non-native speakers to measure the validity of the \textit{Comprehension Quiz} and \textit{Pre-test}.
Most of the participants are bilingual or trilingual (\fgref{demographic_distributions} (B)), and their mother tongues are divided into European and some local languages from Asian countries (such as Japanese, Chinese, Hindi)~(\fgref{demographic_distributions} (C)).


\input{tables/demographic}
\tbref{demographic} shows the demographic data of the participants by groups that we investigated.
We divide demographics according to the following : All participants, people whose \textit{Pre-test} scores were seven or lower (\textit{Pre-test} $\leq$ 7), people whose \textit{Pre-test} scores were above seven (\textit{Pre-test} > 7), English Non-Natives, and English Natives.
The pretest $\leq$ 7 and English Non-Natives were assumed to have almost the same meaning, as well as the pretest > 7 and English natives. However, we noticed that some people who declared that they are native English speakers have lower scores on the \textit{Pre-test} than some of the people who reported that they are non-native English speakers. Therefore, we investigated both to evaluate our system by participants' capabilities and cultural factors.

\subsection{All Participants}

\subsubsection{Comprehension, Engagement, and Readability}
\fgw{all_participants_result_v3}{all_participants_result_v3}{1.0}{Workload and each item on the questionnaire of all participants.}
\textit{Comprehension}, \textit{Engagement}, and \textit{Readability} in \fgref{all_participants_result_v3} show the results of all participants' self-awareness of their comprehension of the video clips, engagement with the clips, and the readability of the subtitles (We asked ``How well did you understand the content of the video?'', ``How engaged were you with the video?'', and ``How would you rate the readability of the video content?'').  

\subsubsection{Workload}
\textit{Mental Demand}, \textit{Physical Demand}, \textit{Temporal Demand}, \textit{Performance}, \textit{Effort}, and \textit{Frustration} in \fgref{all_participants_result_v3} show the participants' self-awareness of the workload during their listening to the news and reading the subtitles.
We attached each statistical test's result in the Appendix (see \tbref{statistical_tests_result_all}, \tbref{statistical_tests_result_leq_7}, \tbref{statistical_tests_result_toefl_g_7}, \tbref{statistical_tests_result_english_non_native}, and \tbref{statistical_tests_result_english_native} ).

\subsubsection{Comprehension Quizzes}
`\textbf{Num Correct}' in \fgref{all_participants_result_v3} shows the number of correct answers to \textit{Comprehension Quiz} by each condition.

\subsection{Other Groups}
\fgw{seven_or_lower_result_v3}{seven_or_lower_result_v3}{1.0}{Workload and each item on the questionnaire of participants whose \textit{Pre-test} scores are seven or lower.}
\fgw{above_7_result_v3}{above_7_result_v3}{1.0}{Workload and each item on the questionnaire of all participants with \textit{Pre-test} scores above seven.}

\fgref{seven_or_lower_result_v3} and \fgref{above_7_result_v3} are the result of the group whose \textit{Pre-test} scores are seven or lower and the group whose \textit{Pre-test} scores are greater than seven.
The tendency of each items are the same as those of all participants in the \textit{Pre-test} $\leq$ 7, but not in the \textit{Pre-test} > 7.
The same can be said for non-native English-speaking people and natives (see \fgref{nonnative_result_v3} and \fgref{native_result_v3} in the appendix).

%% file: tables/demographic.tex
\begin{table*}[]
\centering
\caption{Demographic data of all groups we have investigated.}
\label{tb:demographic}
\begin{tabular}{llccccc}
\hline
\multicolumn{2}{l}{\textbf{Characteristic}} &
  \textbf{All Participants} &
  \textbf{\begin{tabular}[c]{@{}c@{}}Pre-test \\ $\leq$ 7\end{tabular}} &
  \textbf{\begin{tabular}[c]{@{}c@{}}Pre-test \\ \textgreater 7\end{tabular}} &
  \textbf{English Non-Native} &
  \textbf{English Native} \\ \hline
\multirow{3}{*}{Age} &
  mean &
  30.1 &
  31.5 &
  29.3 &
  30.1 &
  30.1 \\
 &
  median &
  27.0 &
  27.0 &
  27.5 &
  26.5 &
  29.0 \\
 &
  range &
  20--67 &
  20--67 &
  20--53 &
  20--67 &
  20--57 \\ \hline
\multirow{3}{*}{Gender} &
  man &
  51 (60.7\,\%) &
  18 (60.0\,\%) &
  33 (61.1\,\%) &
  41 (62.1\,\%) &
  10 (55.6\,\%) \\
 &
  non-binary &
  3 (3.6\,\%) &
  2 (6.7\,\%) &
  1 (1.9\,\%) &
  3 (4.5\,\%) &
  - \\
 &
  woman &
  30 (35.7\,\%) &
  10 (33.3\,\%) &
  20 (37.0\,\%) &
  22 (33.3\,\%) &
  8 (44.4\,\%) \\ \hline
\multirow{2}{*}{\begin{tabular}[c]{@{}l@{}}Pre-test\\ Score\end{tabular}} &
  mean &
  7.42 &
  4.77 &
  8.89 &
  7.05 &
  8.78 \\
 &
  range &
  2--10 &
  2--7 &
  8--10 &
  2--10 &
  7--10 \\ \hline
\multirow{3}{*}{Nationality} &
  1st &
  \begin{tabular}[c]{@{}c@{}}Mexico: \\ 10 (11.9\,\%)\end{tabular} &
  \begin{tabular}[c]{@{}c@{}}Japan: \\ 5 (16.7\,\%)\end{tabular} &
  \begin{tabular}[c]{@{}c@{}}South Africa: \\ 8 (14.8\,\%)\end{tabular} &
  \begin{tabular}[c]{@{}c@{}}Mexico: \\ 10 (15.2\,\%)\end{tabular} &
  \begin{tabular}[c]{@{}c@{}}South Africa: \\ 9 (50.0\,\%)\end{tabular} \\
 &
  2nd &
  \begin{tabular}[c]{@{}c@{}}Portugal: \\ 9 (10.7\,\%)\end{tabular} &
  \begin{tabular}[c]{@{}c@{}}Mexico: \\ 3 (10.0\,\%)\end{tabular} &
  \begin{tabular}[c]{@{}c@{}}Mexico: \\ 7 (13.0\,\%)\end{tabular} &
  \begin{tabular}[c]{@{}c@{}}Portugal: \\ 8 (12.1\,\%)\end{tabular} &
  \begin{tabular}[c]{@{}c@{}}United Kingdom: \\ 5 (27.8\,\%)\end{tabular} \\
 &
  3rd &
  \begin{tabular}[c]{@{}c@{}}South Africa: \\ 9 (10.7\,\%)\end{tabular} &
  \begin{tabular}[c]{@{}c@{}}Portugal: \\ 3 (10.0\,\%)\end{tabular} &
  \begin{tabular}[c]{@{}c@{}}Portugal: \\ 6 (11.1\,\%)\end{tabular} &
  \begin{tabular}[c]{@{}c@{}}Poland: \\ 8 (12.1\,\%)\end{tabular} &
  \begin{tabular}[c]{@{}c@{}}Kenya: \\ 2 (11.1\,\%)\end{tabular} \\ \hline
\multirow{3}{*}{\begin{tabular}[c]{@{}l@{}}Native\\ Language\end{tabular}} &
  1st &
  \begin{tabular}[c]{@{}c@{}}English: \\ 18 (21.4\,\%)\end{tabular} &
  \begin{tabular}[c]{@{}c@{}}Portuguese: \\ 5 (16.7\,\%)\end{tabular} &
  \begin{tabular}[c]{@{}c@{}}English: \\ 17 (31.5\,\%)\end{tabular} &
  \begin{tabular}[c]{@{}c@{}}Spanish: \\ 12 (18.2\,\%)\end{tabular} &
  \begin{tabular}[c]{@{}c@{}}English: \\ 18 (100.0\,\%)\end{tabular} \\
 &
  2nd &
  \begin{tabular}[c]{@{}c@{}}Spanish: \\ 12 (14.3\,\%)\end{tabular} &
  \begin{tabular}[c]{@{}c@{}}Japanese: \\ 5 (16.7\,\%)\end{tabular} &
  \begin{tabular}[c]{@{}c@{}}Spanish: \\ 9 (16.7\,\%)\end{tabular} &
  \begin{tabular}[c]{@{}c@{}}Portuguese: \\ 10 (15.2\,\%)\end{tabular} &
  - \\
 &
  3rd &
  \begin{tabular}[c]{@{}c@{}}Portuguese: \\ 10 (11.9\,\%)\end{tabular} &
  \begin{tabular}[c]{@{}c@{}}Spanish: \\ 3 (10.0\,\%)\end{tabular} &
  \begin{tabular}[c]{@{}c@{}}Chinese: \\ 6 (11.1\,\%)\end{tabular} &
  \begin{tabular}[c]{@{}c@{}}Polish: \\ 8 (12.1\,\%)\end{tabular} &
  - \\ \hline
\end{tabular}
\end{table*}

%% file: 08_discussion.tex
\section{Discussion}

\subsection{Validity of AI-Generated Quizzes}

We used generative AI to create quizzes to assess listening comprehension. Manual corrections were necessary, addressing issues such as unbalanced answer choice, numerical calculation errors, logical inconsistencies, and verbatim extractions from the source text. This approach of using AI for quiz generation could be applied to other areas, such as foreign language learning. Future improvements should focus on these specific areas of weakness.



\subsection{Comprehension Quiz Performance}


The results of \textit{Comprehension Quiz} Performance suggest that there were no significant differences in the average number of correct answers (\textit{Num Correct}) across subtitle conditions, except for the case between \textit{Keyword} and \textit{Dynamik} conditions in the group of people whose \textit{Pre-test} scores are seven or less.
This might be because the quizzes test memory recall more than comprehension, suggesting that while assistive subtitles may aid reading, they may not significantly impact information retention. In fact, some participants indicated this; e.g.,``this didn’t really test listening comprehension as much as memory''.

There were also people who did not notice how the different font sizes related to the importance of the word, although we instructed them in the experiment. 
Although there is a limitation of data clensing on participants who were not engaged in the experiment through crowdsourcing, if we could narrow down participants to whom they fully understood the instruction and committed without skipping it, there is the possibility of a significant difference of the \textit{Comprehension Quiz} results by the conditions.

\subsection{Self-Awareness of the Extent of Comprehension, Readability, and Engagement: Are Non-native Speakers Same as English Dyslexia?}

Regarding self-awareness of comprehension, readability of the texts, and engagement in reading the text, we discuss by group on \textit{Pre-test} scores.

In the group of participants whose \textit{Pre-test} scores are above seven, there were no significant differences.

However, in the group of participants whose \textit{Pre-test} scores are seven or lower, while the \textit{Comprehension} scores showed significant differences, the \textit{Readability} scores and \textit{Engagement} scores did not.
This means that although readability and concentration level while reading were not significantly affected by text conditions, many participants felt that the \textit{Dynamik} subtitles helped improve their comprehension.

This aligns with previous research on dyslexia, suggesting that non-native language readers might have similar characteristics to those with dyslexia when processing text~\cite{rello-etal-2014-keyword}, experiencing similar challenges. If this parallel holds true, methods and services developed for dyslexic individuals could potentially benefit non-native language learners.

As an additional implication of the relativity of these two features, previous studies have shown that other features, such as gaze, have differentiated native / non-native speakers or people with / without dyslexia~\cite{Rello2015_gaze, Fujii2019_subme}. 
In future work, we will investigate the potential correlation of the characteristics of language fluency and dyslexia.

\subsection{NASA--TLX Results}
Here, we also discuss by group on \textit{Pre-test} scores.

As in the same case as \textit{Comprehension}, \textit{Readability} and \textit{Engagement}, in the group of participants whose \textit{Pre-test} scores are above seven, there were no significant differences.

However, in the group of participants whose \textit{Pre-test} scores are seven or lower, significant differences (p < 0.05) were observed in \textit{Performance} and \textit{Effort}. 

This suggests that \textit{Dynamik} facilitates people's sense of performance of reading the sentences, as is the sense of comprehension, and it alleviates people's cognitive workload, in the case when people have relatively low English skills.





\subsection{Applicability to Other Languages}

Although our method could be adapted to other languages, considerations for language-specific characteristics are necessary. For example, Japanese, with its three writing systems, might not benefit as much from size-based emphasis due to the inherent visual cues provided by the kanji characters, which means kanji has more meaning density compared to the alphabet~\cite{kajii_2001}.

\subsection{Font Size Variation}
The size variation in \textit{Dynamik} subtitles might have increased visual stimulation, potentially affecting NASA--TLX results. In this research, we examined pairs of 12\,pt and 18\,pt because they offer optimal readability for a wide audience, including those with dyslexia and the elderly~\cite{10.1145/2461121.2461125, 10.1145/2858036.2858204, 10.1145/634067.634173, bhatia2011effect, 10.1145/2207016.2207048}. However, future studies should explore the ideal ratio of sizes for better results.


\subsection{Potential for Reducing Subtitle Display Area}

By categorizing words into content and function words, we reduce the size of approximately 40\,\% of the words. With function words displayed at 2/3 the size of content words, our method resulted in subtitles occupying about 80\,\% of the original length $(1.0 \times 60\,\% + 0.67 \times 40\,\% \approx 80\,\%)$. This reduction in occupied area could be beneficial, especially for devices with limited display space.

\subsection{Alternative Display Methods}

Based on participant feedback, alternative methods such as using transparency, bold text, color changes, or multilevel size adjustments could be explored in future studies. Our focus on font size was driven by the practical benefit of reducing subtitle area, but other visual cues might prove effective as well. Considering that the tendency of non-native speakers' senses of \textit{Readability}, \textit{Engagement} and \textit{Comprehension} is similar to that of people with dyslexia, other display methods that are useful for people with dyslexia will also work, such as boldness~\cite{rello-etal-2014-keyword}. 

In addition, according to the free-form questionnaire, some people were struggling to use \textit{Dynamik}, saying that they did not notice what the difference in font size meant at first, or it stressed them out.

Also, in this research, we did not explore patterns besides alternative display methods; font size ratio, combination of colors and fonts, and so on. 
More exploration is needed to explore this novice display method.

\subsection{Better Keyword Extraction Method}

Although we used spaCy for our current implementation, some participants noted discrepancies between expected important words and displayed keywords, some participants saying ``important keywords differ depending on each person''. This suggests limitations in classification based solely on parts of speech. 

In this study, we broadly categorized words into function words and content words; however, a more detailed classification could provide deeper insights. For instance, proper nouns may carry greater importance than common nouns. 
Additionally, future improvements could involve light-weight machine learning models, potentially training them to predict the importance of each word (e.g., tf-idf values) for upcoming words.
One idea is that reinforcement learning could be employed to develop AI models that mimic human gaze patterns and cognitive behaviors, serving as a tool for more comprehensive evaluation and training models.

\subsection{System Latency}

Morphological analysis introduces a delay of 0.2--0.3 seconds for space-delimited languages like English and up to 0.5 seconds for languages without spacing. The Markov process used requires processing a chunk of text from the beginning, which adds to the lag. In our real-time system, this resulted in a delay of about 0.5 seconds in displaying speech recognition results. Although this delay was consistent in the recorded videos used for crowd-sourcing, some participants still noted the lag.

%% file: 09_conclusion.tex
\section{Conclusion}

Our study investigated the effectiveness of the \textit{Dynamik} subtitle method, which emphasizes content words and de-emphasizes function words through font size.
The results showed that it significantly improves the user's self-awareness of comprehension and reduces some cognitive load (\textit{Effort} and \textit{Performance}), especially among non-native English speakers with effect sizes of huge (> 0.8).
As global content consumption continues to grow, we believe that research on listening aid methods, such as this research, is crucial to improve cross-cultural communication.

%% file: 10_appendix.tex
\clearpage

\input{tables/video_info}

\input{tables/quiz_info}

\input{tables/statistical_tests_result}

\input{tables/nasa-tlx}

\input{tables/other_metrics}

\fgw{nonnative_result_v3}{nonnative_result_v3}{1.0}{Workload and each item on the questionnaire of non-native English-speaking participants.}

\fgw{native_result_v3}{native_result_v3}{1.0}{Workload and each item on the questionnaire of native English-speaking participants.}

%% file: tables/video_info.tex
\begingroup
\renewcommand{\footnotesize}{\Large}
\renewcommand{\thefootnote}{\arabic{footnote}}
\renewcommand{\thempfootnote}{\arabic{mpfootnote}}

\begin{table*}
\centering
\Large
\begin{minipage}{1.0\linewidth}
\caption{Information of news clips used in the experiment.}
\label{tb:news_clip_info}
\begin{tabular}{lcccclc}
\hline
\textbf{News} &
  \textbf{\begin{tabular}[c]{@{}c@{}}Total \\ words\end{tabular}} &
  \textbf{\begin{tabular}[c]{@{}c@{}}Content \\ words\end{tabular}} &
  \textbf{\begin{tabular}[c]{@{}c@{}}Function \\ words\end{tabular}} &
  \textbf{\begin{tabular}[c]{@{}c@{}}Lexical \\ Density (\%)\end{tabular}} &
  \textbf{News Title} &
  \textbf{Transcription} \\ \hline
1 &
  84 &
  50 &
  34 &
  60 &
  Global wildlife down 60\,\% &
  \footnotemark[1] \\ \hline
2 &
  70 &
  38 &
  32 &
  54 &
  Natural mushroom cloud causes alarm &
  \footnotemark[2] \\ \hline
3 &
  90 &
  53 &
  37 &
  59 &
  Impact of air pollution on children &
  \footnotemark[3] \\ \hline
4 &
  60 &
  35 &
  25 &
  58 &
  LinkedIn holds workplace parents day &
  \footnotemark[4] \\ \hline
5 &
  79 &
  50 &
  29 &
  63 &
  World rankings in kids science skills &
  \footnotemark[5] \\ \hline
6 &
  52 &
  32 &
  20 &
  62 &
  Fake news prompts gunman's raid &
  \footnotemark[6] \\ \hline
\end{tabular}
    \footnotetext[1]{ A new report from the World Wildlife Fund has found that our planet has lost almost 60\,\% of its wildlife in less than half a century. Scientists say the rapid extinction is caused by the loss of habitat that comes from pollution, the exploitation of resources as well as climate change. The report highlights a number of species, elephants, for example, whose numbers have dropped by 5th in just ten years. As for sharks and rays, 1/3 are threatened by overfishing. }
    \footnotetext[2]{ It might look like a sign of nuclear warfare. How would you feel if you saw this big mushroom cloud hanging over your neighborhood? Well, this one appeared in Western Siberia, and according to Russian media, a number of panicky people called emergency services fearing a nuclear attack. I don't blame them. Turns out this formation happens when a thunderstorm. Causes clouds to be blown sideways.}
    \footnotetext[3]{ Now air pollution is a serious global health concern. UNICEF says around 600,000 children under the age of five die every year from pollution related illnesses and also warns that pollutants can permanently damage children's brain development. Around 2 billion children live in places where pollution levels exceed WHO guidelines. And most of the pollution comes from burning fossil fuel and vehicle emissions. But dangers also lie at home. Around 1 billion children live in homes that use wood and coal for cooking and heating. }
    \footnotetext[4]{ Parents across the globe are checking up on their kids at the office right now as part of Linkedin's Bring Your Parents Day. Here you can see pictures from social media showing how the visits are turning out. In a generation where more and more jobs are becoming less traditional and more flexible, LinkedIn says one in three parents cannot describe their kids job. }
    \footnotetext[5]{ Asia is producing some of the world's brightest students. Every four years, 10 and 14 year olds from around the world get ranked in an international math and science study. And Singapore crushes the competition in every category. For instance, among 10 year olds in science, Singapore comes in number one. That's followed by South Korea. Japan in 3rd and then Russia. Hong Kong comes in fifth. Finland is actually the only Europe in top 10. }
    \footnotetext[6]{ And what started out as a conspiracy theory motivated a man to bring a gun to a pizza shop. This one right there in Washington. Police say the gunman apparently believed a fake news story online, and the bogus story falsely claimed that the pizza shop was a site of a child sex ring run by Hillary Clinton and her come.}
\end{minipage}
\end{table*}

\endgroup

%% file: tables/quiz_info.tex
\begin{table*}[]
\centering
\caption{Information of Comprehension Quiz on video number 1.}
\label{tb:quiz_info1}
\begin{tabular}{cllc}
\hline
\begin{tabular}[c]{@{}c@{}}Question \\ Number\end{tabular} &
  Question Statement &
  Choices &
  \begin{tabular}[c]{@{}c@{}}Correct \\ Answer\end{tabular} \\ \hline
1 &
  \begin{tabular}[c]{@{}l@{}}According to the WWF report, \\ by how much has global wildlife decreased in less than half a century?\end{tabular} &
  \begin{tabular}[c]{@{}l@{}}A: About 40\,\%\\ B: About 60\,\%\\ C: About 75\,\%\\ D: About 90\,\%\end{tabular} &
  B \\ \hline
2 &
  \begin{tabular}[c]{@{}l@{}}Over how many years has the elephant population decreased by one-fifth, \\ according to the report?\end{tabular} &
  \begin{tabular}[c]{@{}l@{}}A: 5 years\\ B: 10 years\\ C: 15 years\\ D: 20 years\end{tabular} &
  B \\ \hline
3 &
  What fraction of sharks and rays are threatened by overfishing? &
  \begin{tabular}[c]{@{}l@{}}A: One-fourth\\ B: One-third\\ C: Half\\ D: Two-thirds\end{tabular} &
  B \\ \hline
4 &
  Which organization released the report on global wildlife decline? &
  \begin{tabular}[c]{@{}l@{}}A: United Nations\\ B: World Bank\\ C: World Wildlife Fund\\ D: Greenpeace\end{tabular} &
  C \\ \hline
5 &
  What time frame does the report cover for the wildlife decline? &
  \begin{tabular}[c]{@{}l@{}}A: Less than a quarter century\\ B: Less than half a century\\ C: Less than a century\\ D: More than a century\end{tabular} &
  B \\ \hline
6 &
  What is not the main causes of rapid extinction mentioned in the report? &
  \begin{tabular}[c]{@{}l@{}}A: Natural disasters\\ B: Habitat loss and pollution\\ C: Hunting and poaching\\ D: Climate change\end{tabular} &
  A \\ \hline
7 &
  \begin{tabular}[c]{@{}l@{}}Which specific animal group is mentioned \\ as having lost a fifth of its population in a decade?\end{tabular} &
  \begin{tabular}[c]{@{}l@{}}A: Lions\\ B: Elephants\\ C: Tigers\\ D: Rhinos\end{tabular} &
  B \\ \hline
\end{tabular}
\end{table*}

\begin{table*}[]
\centering
\caption{Information of Comprehension Quiz on video number 2.}
\label{tb:quiz_info2}
\begin{tabular}{cllc}
\hline
\begin{tabular}[c]{@{}c@{}}Question \\ Number\end{tabular} &
  Question Statement &
  Choices &
  \begin{tabular}[c]{@{}c@{}}Correct \\ Answer\end{tabular} \\ \hline
1 &
  Where did this cloud appear? &
  \begin{tabular}[c]{@{}l@{}}A: Western Siberia\\ B: Eastern Siberia\\ C: Northern Siberia\\ D: Southern Siberia\end{tabular} &
  A \\ \hline
2 &
  What did many people fear when they called emergency services? &
  \begin{tabular}[c]{@{}l@{}}A: Forest fire\\ B: Meteor strike\\ C: Volcanic eruption\\ D: Nuclear attack\end{tabular} &
  D \\ \hline
3 &
  What was the actual cause of this cloud formation? &
  \begin{tabular}[c]{@{}l@{}}A: Thunderstorm\\ B: Factory emissions \\ C: Military exercise\\ D: Meteor impact\end{tabular} &
  A \\ \hline
4 &
  How did the news describe people's reaction to seeing the cloud? &
  \begin{tabular}[c]{@{}l@{}}A: Curious\\ B: Excited\\ C: Panicky\\ D: Indifferent\end{tabular} &
  C \\ \hline
5 &
  According to the news what is the shape of the cloud? &
  \begin{tabular}[c]{@{}l@{}}A: Mushroom\\ B: Tornado\\ C: Huge potato\\ D: Thunderstorm\end{tabular} &
  A \\ \hline
6 &
  How did the cloud appear to be formed? &
  \begin{tabular}[c]{@{}l@{}}A: Rising straight up\\ B: Blown sideways\\ C: Spiraling\\ D: Dissipating quickly\end{tabular} &
  B \\ \hline
7 &
  What did the reporter say about people's reaction? &
  \begin{tabular}[c]{@{}l@{}}A: It was an overreaction\\ B: It was understandable\\ C: It was amusing\\ D: It was concerning\end{tabular} &
  B \\ \hline
\end{tabular}
\end{table*}

\begin{table*}[]
\centering
\caption{Information of Comprehension Quiz on video number 3.}
\label{tb:quiz_info3}
\begin{tabular}{cllc}
\hline
\begin{tabular}[c]{@{}c@{}}Question \\ Number\end{tabular} &
  Question Statement &
  Choices &
  \begin{tabular}[c]{@{}c@{}}Correct \\ Answer\end{tabular} \\ \hline
1 &
  \begin{tabular}[c]{@{}l@{}}According to UNICEF, \\ how many children under the age of five \\ die every year from pollution-related illnesses?\end{tabular} &
  \begin{tabular}[c]{@{}l@{}}A: About 200000\\ B: About 400000\\ C: About 600000\\ D: About 800000\end{tabular} &
  C \\ \hline
2 &
  \begin{tabular}[c]{@{}l@{}}How many children live in areas \\ where pollution levels exceed WHO guidelines?\end{tabular} &
  \begin{tabular}[c]{@{}l@{}}A: About 500 million\\ B: About 1 billion\\ C: About 1.5 billion\\ D: About 2 billion\end{tabular} &
  D \\ \hline
3 &
  What is mentioned as a main cause of indoor air pollution? &
  \begin{tabular}[c]{@{}l@{}}A: Keeping pets\\ B: Using pesticides\\ C: Using plastic products\\ D: Using wood and coal \\ for cooking and heating\end{tabular} &
  D \\ \hline
4 &
  \begin{tabular}[c]{@{}l@{}}What long-term effect can pollutants \\ have on children according to UNICEF?\end{tabular} &
  \begin{tabular}[c]{@{}l@{}}A: Worse IQ\\ B: Permanent brain damage\\ C: Less physical growth\\ D: Worse respiratory health\end{tabular} &
  B \\ \hline
5 &
  What are the main sources of pollution mentioned in the news? &
  \begin{tabular}[c]{@{}l@{}}A: Industrial waste\\ B: Fossil fuel burning \\ and vehicle emissions\\ C: Agricultural runoff\\ D: Electronic waste\end{tabular} &
  B \\ \hline
6 &
  \begin{tabular}[c]{@{}l@{}}How many children live in homes \\ using wood and coal for cooking and heating?\end{tabular} &
  \begin{tabular}[c]{@{}l@{}}A: About 500 million\\ B: About 1 billion\\ C: About 1.5 billion\\ D: About 2 billion\end{tabular} &
  B \\ \hline
7 &
  How does the news describe air pollution as a health concern? &
  \begin{tabular}[c]{@{}l@{}}A: Subtle issue\\ B: Serious global concern\\ C: Localized but serious problem\\ D: Improving situation\end{tabular} &
  B \\ \hline
\end{tabular}
\end{table*}

\begin{table*}[]
\centering
\caption{Information of Comprehension Quiz on video number 4.}
\label{tb:quiz_info4}
\begin{tabular}{cllc}
\hline
\begin{tabular}[c]{@{}c@{}}Question \\ Number\end{tabular} &
  Question Statement &
  Choices &
  \begin{tabular}[c]{@{}c@{}}Correct \\ Answer\end{tabular} \\ \hline
1 &
  What is the name of the event held by LinkedIn? &
  \begin{tabular}[c]{@{}l@{}}A: Bring Your Parents Day \\ B: Bring Your Kids to Work Day\\ C: Family Office Day\\ D: LinkedIn Family Event\end{tabular} &
  A \\ \hline
2 &
  \begin{tabular}[c]{@{}l@{}}According to LinkedIn, \\ what fraction of parents cannot describe their children's jobs?\end{tabular} &
  \begin{tabular}[c]{@{}l@{}}A: One-fourth\\ B: One-third\\ C: Half\\ D: Two-thirds\end{tabular} &
  B \\ \hline
3 &
  \begin{tabular}[c]{@{}l@{}}What generational change is mentioned \\ as the background for this event?\end{tabular} &
  \begin{tabular}[c]{@{}l@{}}A: Decreased \\ independence of children\\ B: Weakening of \\ parent-child relationships\\ C: Deterioration of \\ workplace environments \\ D: Diversification \\ and flexibility of jobs\end{tabular} &
  D \\ \hline
4 &
  Where are parents checking up on their kids during this event? &
  \begin{tabular}[c]{@{}l@{}}A: At home\\ B: At school\\ C: At the office\\ D: In public spaces\end{tabular} &
  C \\ \hline
5 &
  How widespread is this event according to the news? &
  \begin{tabular}[c]{@{}l@{}}A: Local to one city\\ B: National event\\ C: Global event\\ D: Limited to tech companies\end{tabular} &
  C \\ \hline
6 &
  How is the event being documented? &
  \begin{tabular}[c]{@{}l@{}}A: Through official reports\\ B: Via social media pictures\\ C: By news reporters\\ D: Through LinkedIn profiles\end{tabular} &
  B \\ \hline
7 &
  How old are the invited people assumed to be? &
  \begin{tabular}[c]{@{}l@{}}A: under 20\\ B: 20s\\ C: 40s to 70s\\ D: above 70s\end{tabular} &
  C \\ \hline
\end{tabular}
\end{table*}

\begin{table*}[]
\centering
\caption{Information of Comprehension Quiz on video number 5.}
\label{tb:quiz_info5}
\begin{tabular}{cllc}
\hline
\begin{tabular}[c]{@{}c@{}}Question \\ Number\end{tabular} &
  Question Statement &
  Choices &
  \begin{tabular}[c]{@{}c@{}}Correct \\ Answer\end{tabular} \\ \hline
1 &
  Which country ranked first in science skills for 10-year-olds? &
  \begin{tabular}[c]{@{}l@{}}A: Singapore\\ B: South Korea\\ C: Japan \\ D: Finland\end{tabular} &
  A \\ \hline
2 &
  How often is this ranking conducted? &
  \begin{tabular}[c]{@{}l@{}}A: Every 2 years\\ B: Every 3 years\\ C: Every 4 years\\ D: Every 5 years\end{tabular} &
  C \\ \hline
3 &
  Which is the only European country to make it into the top 5? &
  \begin{tabular}[c]{@{}l@{}}A: Finland\\ B: France\\ C: Russia\\ D: None of them\end{tabular} &
  D \\ \hline
4 &
  What age groups are included in this international study? &
  \begin{tabular}[c]{@{}l@{}}A: 8 and 12 year olds\\ B: 9 and 13 year olds\\ C: 10 and 14 year olds\\ D: 11 and 15 year olds\end{tabular} &
  C \\ \hline
5 &
  What subjects are included in this international study? &
  \begin{tabular}[c]{@{}l@{}}A: Language and science\\ B: Math and Language\\ C: Math and science\\ D: Math science and language\end{tabular} &
  C \\ \hline
6 &
  How did the news describe Singapore's performance? &
  \begin{tabular}[c]{@{}l@{}}A: Good\\ B: Above average\\ C: Crushes the competition\\ D: Slightly better than others\end{tabular} &
  C \\ \hline
7 &
  What was Russia's ranking in the science category for 10-year-olds? &
  \begin{tabular}[c]{@{}l@{}}A: 1st\\ B: 2nd\\ C: 3rd\\ D: 4th\end{tabular} &
  D \\ \hline
\end{tabular}
\end{table*}

\begin{table*}[]
\centering
\caption{Information of Comprehension Quiz on video number 6.}
\label{tb:quiz_info6}
\begin{tabular}{cllc}
\hline
\begin{tabular}[c]{@{}c@{}}Question \\ Number\end{tabular} &
  Question Statement &
  Choices &
  \begin{tabular}[c]{@{}c@{}}Correct \\ Answer\end{tabular} \\ \hline
1 &
  Where did the man with a gun go? &
  \begin{tabular}[c]{@{}l@{}}A: A pizza shop\\ B: A government agency\\ C: A school \\ D: A bank\end{tabular} &
  A \\ \hline
2 &
  Who was falsely accused in the fake news story? &
  \begin{tabular}[c]{@{}l@{}}A: Donald Trump\\ B: Barack Obama\\ C: Joe Biden\\ D: Hillary Clinton\end{tabular} &
  D \\ \hline
3 &
  What was the fake news story about? &
  \begin{tabular}[c]{@{}l@{}}A: Election fraud\\ B: Child sex trafficking ring\\ C: Money laundering\\ D: Espionage activities\end{tabular} &
  B \\ \hline
4 &
  Where was the pizza shop located? &
  \begin{tabular}[c]{@{}l@{}}A: New York\\ B: Washington\\ C: Chicago\\ D: Los Angeles\end{tabular} &
  B \\ \hline
5 &
  \begin{tabular}[c]{@{}l@{}}According to the news, \\ what motivated the man to bring a gun to the pizza shop?\end{tabular} &
  \begin{tabular}[c]{@{}l@{}}A: Conspiracy theory\\ B: Personal vendetta\\ C: Robbery attempt\\ D: Political protest\end{tabular} &
  A \\ \hline
6 &
  How did the gunman come to believe the fake news story? &
  \begin{tabular}[c]{@{}l@{}}A: Through a newspaper\\ B: On television\\ C: Online\\ D: From a friend\end{tabular} &
  C \\ \hline
7 &
  What did the fake news falsely claim about Hillary Clinton? &
  \begin{tabular}[c]{@{}l@{}}A: She owned the pizza shop\\ B: She ran a child sex ring\\ C: She was hiding there\\ D: She was selling drugs there\end{tabular} &
  B \\ \hline
\end{tabular}
\end{table*}

%% file: tables/statistical_tests_result.tex

\begin{table*}[]
\centering
\Large
\caption{Statistical results by each condition among all participants. For the values of comprehension, engagement, and readability are 0:Not at all -- 7:Perfect on Likert Scale. The value of num correct ranges from 0 to 10. For the values of mental demand, physical demand, temporal demand, performance, effort, and frustration are 0:Perfect -- 7:Worst on Likert Scale. }
\label{tb:statistical_tests_result_all}
\begin{tabular}{llllrrr}
\hline
\textbf{Group} &
  \textbf{Metric} &
  \textbf{Condition 1} &
  \textbf{Condition 2} &
  \textbf{U-statistic} &
  \textbf{p-value} &
  \textbf{Cohen's d} \\ \hline
\multirow{30}{*}{All Participants} &
  \multirow{3}{*}{comprehension} &
  normal &
  keyword &
  3286.00 &
  0.4306 &
  -0.16 \\
 &                                   & normal  & dynamik & 4334.00 & \textbf{0.0077} & 0.41  \\
 &                                   & keyword & dynamik & 4498.00 & \textbf{0.0014} & 0.55  \\ \cline{2-7} 
 & \multirow{3}{*}{engagement}       & normal  & keyword & 3093.00 & 0.1548          & -0.25 \\
 &                                   & normal  & dynamik & 3923.00 & 0.1957          & 0.20  \\
 &                                   & keyword & dynamik & 4348.00 & \textbf{0.0072} & 0.44  \\ \cline{2-7} 
 & \multirow{3}{*}{readability}      & normal  & keyword & 2351.50 & \textbf{0.0001} & -0.64 \\
 &                                   & normal  & dynamik & 3417.00 & 0.7192          & -0.10 \\
 &                                   & keyword & dynamik & 4528.00 & \textbf{0.0013} & 0.51  \\ \cline{2-7} 
 & \multirow{3}{*}{num\_correct}     & normal  & keyword & 3551.00 & 0.9418          & 0.08  \\
 &                                   & normal  & dynamik & 2993.00 & 0.0811          & -0.23 \\
 &                                   & keyword & dynamik & 3028.50 & 0.1041          & -0.29 \\ \cline{2-7} 
 & \multirow{3}{*}{mental\_demand}   & normal  & keyword & 3329.00 & 0.5210          & -0.08 \\
 &                                   & normal  & dynamik & 3683.50 & 0.6168          & 0.13  \\
 &                                   & keyword & dynamik & 3827.50 & 0.3331          & 0.21  \\ \cline{2-7} 
 & \multirow{3}{*}{physical\_demand} & normal  & keyword & 3019.00 & 0.0989          & -0.26 \\
 &                                   & normal  & dynamik & 3567.50 & 0.8987          & 0.02  \\
 &                                   & keyword & dynamik & 4077.00 & 0.0742          & 0.28  \\ \cline{2-7} 
 & \multirow{3}{*}{temporal\_demand} & normal  & keyword & 2964.50 & 0.0686          & -0.27 \\
 &                                   & normal  & dynamik & 3277.50 & 0.4190          & -0.14 \\
 &                                   & keyword & dynamik & 3801.00 & 0.3793          & 0.13  \\ \cline{2-7} 
 & \multirow{3}{*}{performance}      & normal  & keyword & 3642.00 & 0.7134          & 0.06  \\
 &                                   & normal  & dynamik & 4409.00 & \textbf{0.0044} & 0.44  \\
 &                                   & keyword & dynamik & 4278.00 & \textbf{0.0155} & 0.37  \\ \cline{2-7} 
 & \multirow{3}{*}{effort}           & normal  & keyword & 3320.50 & 0.5007          & -0.11 \\
 &                                   & normal  & dynamik & 4250.50 & \textbf{0.0195} & 0.39  \\
 &                                   & keyword & dynamik & 4427.50 & \textbf{0.0037} & 0.49  \\ \cline{2-7} 
 & \multirow{3}{*}{frustration}      & normal  & keyword & 2887.50 & \textbf{0.0390} & -0.32 \\
 &                                   & normal  & dynamik & 3478.00 & 0.8732          & -0.03 \\
 &                                   & keyword & dynamik & 4094.50 & 0.0682          & 0.29  \\ \hline
\end{tabular}
\end{table*}

\begin{table*}[]
\centering
\Large
\caption{Statistical results by each condition among participants whose Pre-test scores are 7 or lower. For the values of comprehension, engagement, and readability are 0:Not at all -- 7:Perfect on Likert Scale. The value of num correct ranges from 0 to 10. For the values of mental demand, physical demand, temporal demand, performance, effort, and frustration are 0:Perfect -- 7:Worst on Likert Scale. }
\label{tb:statistical_tests_result_leq_7}
\begin{tabular}{llllrrr}
\hline
\textbf{Group} &
  \textbf{Metric} &
  \textbf{Condition 1} &
  \textbf{Condition 2} &
  \textbf{U-statistic} &
  \textbf{p-value} &
  \textbf{Cohen's d} \\ \hline
\multirow{30}{*}{Pre-test score $\leq$ 7} &
  \multirow{3}{*}{comprehension} &
  normal &
  keyword &
  383.00 &
  0.3137 &
  -0.31 \\
 &                                   & normal  & dynamik & 626.00  & \textbf{0.0074} & 0.67  \\
 &                                   & keyword & dynamik & 653.00  & \textbf{0.0022} & 0.87  \\ \cline{2-7} 
 & \multirow{3}{*}{engagement}       & normal  & keyword & 337.00  & 0.0888          & -0.50 \\
 &                                   & normal  & dynamik & 575.00  & 0.0597          & 0.45  \\
 &                                   & keyword & dynamik & 669.00  & \textbf{0.0010} & 0.91  \\ \cline{2-7} 
 & \multirow{3}{*}{readability}      & normal  & keyword & 224.00  & \textbf{0.0007} & -0.98 \\
 &                                   & normal  & dynamik & 517.00  & 0.3156          & 0.23  \\
 &                                   & keyword & dynamik & 711.50  & \textbf{0.0001} & 1.19  \\ \cline{2-7} 
 & \multirow{3}{*}{num\_correct}     & normal  & keyword & 564.50  & 0.0870          & 0.46  \\
 &                                   & normal  & dynamik & 365.50  & 0.2023          & -0.32 \\
 &                                   & keyword & dynamik & 268.50  & \textbf{0.0064} & -0.74 \\ \cline{2-7} 
 & \multirow{3}{*}{mental\_demand}   & normal  & keyword & 337.50  & 0.0862          & -0.35 \\
 &                                   & normal  & dynamik & 522.00  & 0.2792          & 0.41  \\
 &                                   & keyword & dynamik & 599.00  & \textbf{0.0250} & 0.67  \\ \cline{2-7} 
 & \multirow{3}{*}{physical\_demand} & normal  & keyword & 408.00  & 0.5216          & -0.19 \\
 &                                   & normal  & dynamik & 481.50  & 0.6340          & 0.07  \\
 &                                   & keyword & dynamik & 521.00  & 0.2767          & 0.24  \\ \cline{2-7} 
 & \multirow{3}{*}{temporal\_demand} & normal  & keyword & 325.00  & 0.0565          & -0.48 \\
 &                                   & normal  & dynamik & 415.50  & 0.6042          & -0.17 \\
 &                                   & keyword & dynamik & 525.50  & 0.2556          & 0.28  \\ \cline{2-7} 
 & \multirow{3}{*}{performance}      & normal  & keyword & 367.50  & 0.2125          & -0.28 \\
 &                                   & normal  & dynamik & 604.50  & \textbf{0.0202} & 0.66  \\
 &                                   & keyword & dynamik & 642.00  & \textbf{0.0040} & 0.80  \\ \cline{2-7} 
 & \multirow{3}{*}{effort}           & normal  & keyword & 356.50  & 0.1585          & -0.33 \\
 &                                   & normal  & dynamik & 610.00  & \textbf{0.0157} & 0.72  \\
 &                                   & keyword & dynamik & 678.50  & \textbf{0.0006} & 1.01  \\ \cline{2-7} 
 & \multirow{3}{*}{frustration}      & normal  & keyword & 254.00  & \textbf{0.0032} & -0.71 \\
 &                                   & normal  & dynamik & 480.00  & 0.6572          & 0.16  \\
 &                                   & keyword & dynamik & 632.50  & \textbf{0.0061} & 0.77  \\ \hline
\end{tabular}
\end{table*}

\begin{table*}[]
\centering
\Large
\caption{Statistical results by each condition among participants whose Pre-test scores are higher than seven. For the values of comprehension, engagement, and readability are 0:Not at all -- 7:Perfect on Likert Scale. The value of num correct ranges from 0 to 10. For the values of mental demand, physical demand, temporal demand, performance, effort, and frustration are 0:Perfect -- 7:Worst on Likert Scale. }
\label{tb:statistical_tests_result_toefl_g_7}
\begin{tabular}{llllrrr}
\hline
\textbf{Group} &
  \textbf{Metric} &
  \textbf{Condition 1} &
  \textbf{Condition 2} &
  \textbf{U-statistic} &
  \textbf{p-value} &
  \textbf{Cohen's d} \\ \hline
\multirow{30}{*}{Pre-test Score \textgreater 7} &
  \multirow{3}{*}{comprehension} &
  normal &
  keyword &
  1387.50 &
  0.6556 &
  -0.08 \\
 &                                   & normal  & dynamik & 1633.00 & 0.2585          & 0.28  \\
 &                                   & keyword & dynamik & 1703.50 & 0.1141          & 0.37  \\ \cline{2-7} 
 & \multirow{3}{*}{engagement}       & normal  & keyword & 1365.50 & 0.5513          & -0.13 \\
 &                                   & normal  & dynamik & 1459.00 & 0.9974          & 0.05  \\
 &                                   & keyword & dynamik & 1557.00 & 0.5255          & 0.18  \\ \cline{2-7} 
 & \multirow{3}{*}{readability}      & normal  & keyword & 1098.50 & \textbf{0.0233} & -0.49 \\
 &                                   & normal  & dynamik & 1252.00 & 0.1935          & -0.30 \\
 &                                   & keyword & dynamik & 1603.00 & 0.3642          & 0.18  \\ \cline{2-7} 
 & \multirow{3}{*}{num\_correct}     & normal  & keyword & 1265.50 & 0.2245          & -0.19 \\
 &                                   & normal  & dynamik & 1269.50 & 0.2340          & -0.17 \\
 &                                   & keyword & dynamik & 1455.50 & 0.9899          & 0.01  \\ \cline{2-7} 
 & \multirow{3}{*}{mental\_demand}   & normal  & keyword & 1500.50 & 0.7927          & 0.03  \\
 &                                   & normal  & dynamik & 1467.50 & 0.9550          & 0.01  \\
 &                                   & keyword & dynamik & 1394.00 & 0.6902          & -0.02 \\ \cline{2-7} 
 & \multirow{3}{*}{physical\_demand} & normal  & keyword & 1219.00 & 0.1320          & -0.29 \\
 &                                   & normal  & dynamik & 1414.00 & 0.7825          & 0.00  \\
 &                                   & keyword & dynamik & 1662.50 & 0.1987          & 0.30  \\ \cline{2-7} 
 & \multirow{3}{*}{temporal\_demand} & normal  & keyword & 1295.00 & 0.3102          & -0.19 \\
 &                                   & normal  & dynamik & 1358.00 & 0.5345          & -0.13 \\
 &                                   & keyword & dynamik & 1512.00 & 0.7383          & 0.07  \\ \cline{2-7} 
 & \multirow{3}{*}{performance}      & normal  & keyword & 1613.00 & 0.3328          & 0.23  \\
 &                                   & normal  & dynamik & 1712.00 & 0.1115          & 0.34  \\
 &                                   & keyword & dynamik & 1576.00 & 0.4597          & 0.13  \\ \cline{2-7} 
 & \multirow{3}{*}{effort}           & normal  & keyword & 1490.00 & 0.8426          & 0.00  \\
 &                                   & normal  & dynamik & 1642.00 & 0.2502          & 0.24  \\
 &                                   & keyword & dynamik & 1624.50 & 0.2989          & 0.24  \\ \cline{2-7} 
 & \multirow{3}{*}{frustration}      & normal  & keyword & 1352.50 & 0.5113          & -0.15 \\
 &                                   & normal  & dynamik & 1365.50 & 0.5653          & -0.13 \\
 &                                   & keyword & dynamik & 1475.50 & 0.9154          & 0.02  \\ \hline
\end{tabular}
\end{table*}

\begin{table*}[]
\centering
\Large
\caption{Statistical results by each condition among non-native English-speaking participants. For the values of comprehension, engagement, and readability are 0:Not at all -- 7:Perfect on Likert Scale. The value of num correct ranges from 0 to 10. For the values of mental demand, physical demand, temporal demand, performance, effort, and frustration are 0:Perfect -- 7:Worst on Likert Scale.}
\label{tb:statistical_tests_result_english_non_native}
\begin{tabular}{llllrrr}
\hline
\textbf{Group} &
  \textbf{Metric} &
  \textbf{Condition 1} &
  \textbf{Condition 2} &
  \textbf{U-statistic} &
  \textbf{p-value} &
  \textbf{Cohen's d} \\ \hline
\multirow{30}{*}{English Non-Native} &
  \multirow{3}{*}{comprehension} &
  normal &
  keyword &
  2150.00 &
  0.4698 &
  -0.15 \\
 &                                   & normal  & dynamik & 2961.00 & \textbf{0.0033} & 0.52  \\
 &                                   & keyword & dynamik & 3067.50 & \textbf{0.0007} & 0.65  \\ \cline{2-7} 
 & \multirow{3}{*}{engagement}       & normal  & keyword & 2036.50 & 0.2183          & -0.25 \\
 &                                   & normal  & dynamik & 2631.50 & 0.1529          & 0.25  \\
 &                                   & keyword & dynamik & 2909.50 & \textbf{0.0074} & 0.50  \\ \cline{2-7} 
 & \multirow{3}{*}{readability}      & normal  & keyword & 1464.00 & \textbf{0.0002} & -0.69 \\
 &                                   & normal  & dynamik & 2396.00 & 0.7084          & 0.02  \\
 &                                   & keyword & dynamik & 3166.00 & \textbf{0.0002} & 0.67  \\ \cline{2-7} 
 & \multirow{3}{*}{num\_correct}     & normal  & keyword & 2394.50 & 0.7154          & 0.12  \\
 &                                   & normal  & dynamik & 2060.00 & 0.2614          & -0.15 \\
 &                                   & keyword & dynamik & 2008.50 & 0.1773          & -0.25 \\ \cline{2-7} 
 & \multirow{3}{*}{mental\_demand}   & normal  & keyword & 2280.00 & 0.8889          & 0.02  \\
 &                                   & normal  & dynamik & 2578.00 & 0.2405          & 0.27  \\
 &                                   & keyword & dynamik & 2569.00 & 0.2549          & 0.25  \\ \cline{2-7} 
 & \multirow{3}{*}{physical\_demand} & normal  & keyword & 2021.50 & 0.1949          & -0.23 \\
 &                                   & normal  & dynamik & 2305.00 & 0.9767          & 0.01  \\
 &                                   & keyword & dynamik & 2604.50 & 0.1893          & 0.24  \\ \cline{2-7} 
 & \multirow{3}{*}{temporal\_demand} & normal  & keyword & 1968.50 & 0.1275          & -0.27 \\
 &                                   & normal  & dynamik & 2077.00 & 0.2974          & -0.20 \\
 &                                   & keyword & dynamik & 2406.50 & 0.6774          & 0.07  \\ \cline{2-7} 
 & \multirow{3}{*}{performance}      & normal  & keyword & 2486.50 & 0.4390          & 0.13  \\
 &                                   & normal  & dynamik & 2921.00 & \textbf{0.0070} & 0.47  \\
 &                                   & keyword & dynamik & 2735.00 & \textbf{0.0612} & 0.32  \\ \cline{2-7} 
 & \multirow{3}{*}{effort}           & normal  & keyword & 2049.00 & 0.2415          & -0.20 \\
 &                                   & normal  & dynamik & 2824.00 & \textbf{0.0231} & 0.41  \\
 &                                   & keyword & dynamik & 3034.00 & \textbf{0.0014} & 0.59  \\ \cline{2-7} 
 & \multirow{3}{*}{frustration}      & normal  & keyword & 1901.50 & 0.0698          & -0.31 \\
 &                                   & normal  & dynamik & 2314.00 & 0.9947          & 0.00  \\
 &                                   & keyword & dynamik & 2699.00 & 0.0876          & 0.30  \\ \hline
\end{tabular}
\end{table*}

\begin{table*}[]
\centering
\Large
\caption{Statistical results by each condition among native English-speaking participants. For the values of comprehension, engagement, and readability are 0:Not at all -- 7:Perfect on Likert Scale. The value of num correct ranges from 0 to 10. For the values of mental demand, physical demand, temporal demand, performance, effort, and frustration are 0:Perfect -- 7:Worst on Likert Scale.}
\label{tb:statistical_tests_result_english_native}
\begin{tabular}{llllrrr}
\hline
\textbf{Group} &
  \textbf{Metric} &
  \textbf{Condition 1} &
  \textbf{Condition 2} &
  \textbf{U-statistic} &
  \textbf{p-value} &
  \textbf{Cohen's d} \\ \hline
\multirow{30}{*}{English Native} &
  \multirow{3}{*}{comprehension} &
  normal &
  keyword &
  120.00 &
  0.7663 &
  -0.24 \\
 &                                   & normal  & dynamik & 127.00  & 0.9839          & -0.05 \\
 &                                   & keyword & dynamik & 133.50  & 0.8418          & 0.18  \\ \cline{2-7} 
 & \multirow{3}{*}{engagement}       & normal  & keyword & 115.00  & 0.6093          & -0.33 \\
 &                                   & normal  & dynamik & 135.00  & 0.7865          & -0.07 \\
 &                                   & keyword & dynamik & 144.50  & 0.5225          & 0.22  \\ \cline{2-7} 
 & \multirow{3}{*}{readability}      & normal  & keyword & 100.00  & 0.2816          & -0.47 \\
 &                                   & normal  & dynamik & 87.50   & 0.1184          & -0.63 \\
 &                                   & keyword & dynamik & 118.50  & 0.7299          & -0.07 \\ \cline{2-7} 
 & \multirow{3}{*}{num\_correct}     & normal  & keyword & 112.00  & 0.5468          & -0.07 \\
 &                                   & normal  & dynamik & 85.00   & 0.0944          & -0.66 \\
 &                                   & keyword & dynamik & 106.00  & 0.3933          & -0.48 \\ \cline{2-7} 
 & \multirow{3}{*}{mental\_demand}   & normal  & keyword & 92.00   & 0.1660          & -0.47 \\
 &                                   & normal  & dynamik & 96.00   & 0.2169          & -0.40 \\
 &                                   & keyword & dynamik & 127.00  & 0.9843          & 0.04  \\ \cline{2-7} 
 & \multirow{3}{*}{physical\_demand} & normal  & keyword & 99.00   & 0.2742          & -0.39 \\
 &                                   & normal  & dynamik & 135.00  & 0.8013          & 0.07  \\
 &                                   & keyword & dynamik & 160.50  & 0.2192          & 0.44  \\ \cline{2-7} 
 & \multirow{3}{*}{temporal\_demand} & normal  & keyword & 107.00  & 0.4198          & -0.34 \\
 &                                   & normal  & dynamik & 134.00  & 0.8296          & 0.08  \\
 &                                   & keyword & dynamik & 154.00  & 0.3162          & 0.42  \\ \cline{2-7} 
 & \multirow{3}{*}{performance}      & normal  & keyword & 108.00  & 0.4547          & -0.19 \\
 &                                   & normal  & dynamik & 147.00  & 0.4740          & 0.37  \\
 &                                   & keyword & dynamik & 171.00  & 0.1020          & 0.63  \\ \cline{2-7} 
 & \multirow{3}{*}{effort}           & normal  & keyword & 149.50  & 0.4123          & 0.25  \\
 &                                   & normal  & dynamik & 150.50  & 0.3891          & 0.38  \\
 &                                   & keyword & dynamik & 131.50  & 0.9077          & 0.13  \\ \cline{2-7} 
 & \multirow{3}{*}{frustration}      & normal  & keyword & 104.00  & 0.3485          & -0.39 \\
 &                                   & normal  & dynamik & 114.50  & 0.6091          & -0.15 \\
 &                                   & keyword & dynamik & 141.00  & 0.6240          & 0.25  \\ \hline
\end{tabular}
\end{table*}

%% file: tables/nasa-tlx.tex
\begin{table*}[]
\centering

\caption{NASA-TLX Scores by Participant Group (0:Low -- 7:High)}
\label{tb:nasa_tlx}
\begin{tabular}{llrrrrr}
\hline
\multirow{2}{*}{\textbf{Metric}} & \multirow{2}{*}{\textbf{Statistic}} & \multicolumn{5}{c}{\textbf{Participant Group}}                                                                            \\ \cline{3-7} 
                        &                            & \textbf{All Participants} & \textbf{Pre-test Score $\leq$ 7} & \textbf{Pre-test Score \textgreater 7} & \textbf{English Non-Native} & \textbf{English Native} \\ \hline
\multirow{3}{*}{Mental Demand}   & Mean   & 3.11     & 2.98     & 3.18     & 3.05     & 3.35     \\
                                 & Median & 3.0      & 3.0      & 4.0      & 3.0      & 4.0      \\
                                 & Range  & 0.0--6.0 & 0.0--6.0 & 0.0--6.0 & 0.0--6.0 & 0.0--5.0 \\ \hline
\multirow{3}{*}{Physical Demand} & Mean   & 1.79     & 1.94     & 1.70     & 1.71     & 2.12     \\
                                 & Median & 1.0      & 2.0      & 1.0      & 1.0      & 2.0      \\
                                 & Range  & 0.0--5.0 & 0.0--5.0 & 0.0--5.0 & 0.0--5.0 & 0.0--5.0 \\ \hline
\multirow{3}{*}{Temporal Demand} & Mean   & 2.89     & 2.92     & 2.88     & 2.75     & 3.50     \\
                                 & Median & 3.0      & 3.0      & 3.0      & 3.0      & 4.0      \\
                                 & Range  & 0.0--6.0 & 0.0--6.0 & 0.0--6.0 & 0.0--6.0 & 0.0--5.0 \\ \hline
\multirow{3}{*}{Performance}     & Mean   & 2.59     & 2.69     & 2.54     & 2.71     & 2.10     \\
                                 & Median & 2.0      & 3.0      & 2.0      & 3.0      & 2.0      \\
                                 & Range  & 0.0--6.0 & 0.0--6.0 & 0.0--6.0 & 0.0--6.0 & 0.0--6.0 \\ \hline
\multirow{3}{*}{Effort}          & Mean   & 2.90     & 2.72     & 2.99     & 2.73     & 3.60     \\
                                 & Median & 3.0      & 3.0      & 3.0      & 3.0      & 4.0      \\
                                 & Range  & 0.0--6.0 & 0.0--6.0 & 0.0--6.0 & 0.0--6.0 & 0.0--6.0 \\ \hline
\multirow{3}{*}{Frustration}     & Mean   & 2.59     & 2.93     & 2.40     & 2.76     & 1.88     \\
                                 & Median & 3.0      & 3.0      & 3.0      & 3.0      & 2.5      \\
                                 & Range  & 0.0--6.0 & 0.0--6.0 & 0.0--6.0 & 0.0--6.0 & 0.0--6.0 \\ \hline
\end{tabular}
\end{table*}

%% file: tables/other_metrics.tex
\begin{table*}[htbp]
\centering
\Large
\caption{Other Metrics by Participant Group (As for Comprehension, Engagement, Readability, 0:Not at all -- 7:Perfect on Likert Scale. Num Correct ranges from 0 to 10.)}
\label{tb:other_metrics}
\begin{tabular}{llrrrrr}
\hline
\multirow{2}{*}{\textbf{Metric}} & \multirow{2}{*}{\textbf{Statistic}} & \multicolumn{5}{c}{\textbf{Participant Group}}                                                          \\ \cline{3-7} 
                        &                            & \textbf{All Participants} & \textbf{TOEFL $\leq$ 7} & \textbf{TOEFL \textgreater 7} & \textbf{English Non-Native} & \textbf{English Native} \\ \hline
\multirow{3}{*}{Num Correct}   & Mean   & 3.94     & 3.51     & 4.18     & 3.85     & 4.34     \\
                               & Median & 4.0      & 3.0      & 4.0      & 4.0      & 5.0      \\
                               & Range  & 0.0--7.0 & 0.0--7.0 & 0.0--7.0 & 0.0--7.0 & 1.0--7.0 \\ \hline
\multirow{3}{*}{Comprehension} & Mean   & 3.71     & 3.35     & 3.91     & 3.59     & 4.23     \\
                               & Median & 4.0      & 3.0      & 4.0      & 4.0      & 4.0      \\
                               & Range  & 0.0--6.0 & 0.0--6.0 & 0.0--6.0 & 0.0--6.0 & 0.0--6.0 \\ \hline
\multirow{3}{*}{Engagement}    & Mean   & 3.57     & 3.21     & 3.77     & 3.46     & 4.03     \\
                               & Median & 4.0      & 3.0      & 4.0      & 4.0      & 4.0      \\
                               & Range  & 0.0--6.0 & 0.0--6.0 & 0.0--6.0 & 0.0--6.0 & 0.0--6.0 \\ \hline
\multirow{3}{*}{Readability}   & Mean   & 3.88     & 3.52     & 4.09     & 3.78     & 4.28     \\
                               & Median & 4.0      & 4.0      & 4.0      & 4.0      & 4.0      \\
                               & Range  & 0.0--6.0 & 0.0--6.0 & 0.0--6.0 & 0.0--6.0 & 1.0--6.0 \\ \hline
\end{tabular}
\end{table*}